\documentclass[conference,pdftex]{IEEEtran}

\usepackage{todonotes}
\pdfoutput=1
    \usepackage{graphicx}
    \graphicspath{{../pdf/}{../jpeg/}}
    \DeclareGraphicsExtensions{.pdf,.jpeg,.png}
    \usepackage{lipsum}
    \usepackage[cmex10]{amsmath}
    \usepackage{amsthm, amssymb, amsfonts}
	\usepackage{mathabx}
	\usepackage{algorithmic}
	\usepackage{array}
	\usepackage{mdwmath}
	\usepackage{mdwtab}
	\usepackage{eqparbox}
	\usepackage{url}
    \usepackage{subfigure}
    
    \usepackage{breakurl}
\usepackage[
    linkcolor=black,
    urlcolor=black,
    citecolor=teal,
    hidelinks,
    pdftitle={A clustering aggregation algorithm on neutral-atoms and annealing quantum processors},
    pdfauthor={Riccardo Scotti, Gabriella Bettonte, Antonio Costantini, Sara Marzella, Daniele Ottaviani, Stefano Lodi}
]{hyperref}

\let\labelindent\relax
\usepackage{enumitem}
\usepackage{braket}

\DeclareMathOperator*{\argmax}{arg\,max}

\begin{document}

\title{\LARGE A clustering aggregation algorithm
on neutral-atoms and annealing
quantum processors}

 \author{ \authorblockA{}}

\author{\authorblockN{Riccardo Scotti\authorrefmark{1}, Gabriella Bettonte\authorrefmark{2}, Antonio Costantini\authorrefmark{2}, Sara Marzella\authorrefmark{2}, Daniele Ottaviani\authorrefmark{2}, Stefano Lodi\authorrefmark{1} }
\authorblockA{\authorrefmark{1}University of Bologna, Bologna, Italy}
\authorblockA{\authorrefmark{2}CINECA, Casalecchio di Reno, Italy\\a.costantini@cineca.it}}

\maketitle

\begin{abstract}
This work presents a hybrid quantum-classical algorithm to perform clustering aggregation, designed for neutral-atoms quantum computers and quantum annealers. Clustering aggregation is a technique that mitigates the weaknesses of clustering algorithms, an important class of data science methods for partitioning datasets, and is widely employed in many real-world applications. By expressing the clustering aggregation problem instances as a Maximum Independent Set (MIS) problem and as a Quadratic Unconstrained Binary Optimization (QUBO) problem, it was possible to solve them by leveraging the potential of Pasqal's Fresnel (neutral-atoms processor) and D-Wave's Advantage QPU (quantum annealer). Additionally, the designed clustering aggregation algorithm was first validated on a Fresnel emulator based on QuTiP and later on an emulator of the same machine based on tensor networks, provided by Pasqal. The results revealed technical limitations, such as the difficulty of adding additional constraints on the employed neutral-atoms platform and the need for better metrics to measure the quality of the produced clusterings. However, this work represents a step towards a benchmark to compare two different machines: a quantum annealer and a neutral-atom quantum computer. Moreover, findings suggest promising potential for future advancements in hybrid quantum-classical pipelines, although further improvements are needed in both quantum and classical components.
\end{abstract}

\IEEEoverridecommandlockouts
\begin{keywords}
Quantum Computing, Clustering aggregation, QUBO, MIS, Quantum Annelaer, Neutral Atom quantum computer, D-wave, Pasqal
\end{keywords}

\IEEEpeerreviewmaketitle


\section{Introduction}

Quantum computing is a promising branch of computer science that offers a completely new paradigm of computation and a different perspective on how information is encoded and manipulated. 
Different areas of research have been able to produce remarkable discoveries, demonstrating that quantum technologies have the potential to revolutionize numerous fields, such as cryptography \cite{Shor1997}, machine learning \cite{Sood2023}, materials science \cite{Alexeev2024} and many more.

Despite the numerous groundbreaking theoretical results, quantum utility, or the use of quantum computers to solve practical problems, is still far from reality, mainly because of technical limitations in the construction of quantum machines \cite{Herrmann2023}. At the same time, quantum technologies in industry are moving their first steps, with research and supercomputing centers beginning to include them as part of their environment and large investments coming from both private and public organizations. For example, the European High Performance Computing Joint Undertaking (EuroHPC JU), a major initiative by the European Union aimed at developing and supporting a large supercomputing infrastructure across Europe, has recognized the potential of quantum computing, and part of its strategic agenda is the integration of such technologies in the supercomputing ecosystems of the continent \cite{EuroHPC}. 
Instead of replacing classical computers, quantum technologies are now viewed as potential components that can complement classical systems by solving specific types of problems more efficiently \cite{Hamilton2020}. Researchers are currently exploring the idea that quantum computers can serve as accelerators when integrated into heterogeneous HPC clusters. The intuition behind this idea is that, similar to GPUs that have been designed for other purposes now become critical to accelerate computation in an HPC environment, quantum computers can share the same destiny. 
Given that the current technological landscape of quantum computing is uncertain and fragmented, with many available typologies of quantum machines, but no clear superior one, it is worth exploring all of them. Most likely in the future 
researchers and industries will choose a machine that is better suited to address the problem according to hardware features. 


 
 This study focuses on the development of an algorithm to perform clustering aggregation, a data analysis task, designed specifically for neutral atoms, and annealing quantum processors. Clustering is a ubiquitous data science technique with important applications in critical fields such as image processing, marketing, and energy distribution. However, many clustering algorithms present weaknesses when applied to certain data distributions, which can be mitigated by clustering aggregation, a method that produces a unique and robust solution from the results of multiple clustering algorithms. While there are many reasons to focus on this problem, such as its industrial relevance and the possibility of improving its scalability, we claim that clustering aggregation is an ideal candidate use case for testing the efficient design of a hybrid classical quantum algorithm. We considered hybrid pipelines that employ neutral atoms and annealing processors. The clustering aggregation problem can be reduced to a Maximum Independent Set problem, an optimization problem that is naturally solved by neutral-atom machines, and to QUBO problem, a class of optimization problems that are naturally tackled by quantum annealers. 
 
 We tested the newly designed algorithm on real quantum hardware, namely Pasqal's neutral-atoms based processor Fresnel, and D-Wave's quantum annealer Advantage (both via cloud). Not only was it possible to assess the current effectiveness of such machines within a hybrid pipeline, but also to measure performance and compare the strengths and weaknesses of the technologies from a practical point of view, which is particularly relevant, considering that the benchmarking of quantum computers is still an open research area with no single and well-established methodology, especially for non-gate-based processors \cite{Barbaresco2024}. 
 Different benchmark suites for quantum computers have been proposed (for example, a quantum version of LINPACK, a benchmark algorithm used to rank supercomputers in the TOP500 list \cite{Linpack2020, Dong2021}); however, because of the heterogeneity of quantum machines, the metrics they offer do not clearly reflect how processors perform on various classes of algorithms and are strictly related to the underlying technology. For these reasons, better alternatives for assessing the performance of quantum computers are application-oriented benchmarks, in which the quality judgment of the machine is given based on its practical capabilities at reaching a quantum advantage for specific tasks, mainly NP-hard problems such as the Maximum Independent Set (MIS) and Maximum-Weight Independent Set (MWIS). The main goals of this work are to design a quantum algorithm to perform clustering aggregation, test it on real hardware, and compare results across the chosen technologies (i.e., neutral atom quantum computers and quantum annealers).


\section{Background}

In this section, we introduce the basic concepts about clustering aggregation, quantum annealing, and neutral-atom quantum computers. These notions are critical for delving into the work presented in this study.


\subsection{Clustering aggregation}

Clustering is a widely employed machine learning method, with use cases ranging from image processing to finance. However, different clustering algorithms often yield divergent results due to differences in their underlying assumptions, biases or sensitivities to the input data. This variability presents a challenge when attempting to derive reliable clustering results by using a single algorithm. 

Clustering aggregation, or clustering ensemble, is a framework developed specifically to address this issue by combining multiple clustering solutions into a single solution, thereby improving the overall quality of the solution \cite{strehl2002cluster}. 

Because clustering aggregation is an NP-hard problem (see Section \ref{sec:aggregation_graph} for additional details), classical algorithms for this task often present scalability issues as the problem size increases, requiring many trade-offs in terms of quality or execution times \cite{Gionis2007}.

Therefore, the exploration of alternative technologies, such as quantum computers, to address this task is a direction worth following, as it may lead to algorithms with improved accuracy (i.e., producing solutions of higher quality).

\subsection{Quantum annealing}


Quantum annealing is a meta-heuristic optimization technique that exploits the principles of quantum mechanics, such as quantum tunneling and adiabatic theorem, to find the global minimum of a given objective function \cite{Rajak2023, Johnson2011}. It is mainly used as a solver for combinatorial optimization problems in lieu of analogous classical algorithms, such as simulated annealing. 

The key idea in quantum annealing is to encode the problem function into the energy states of a quantum system such that the optimal solution corresponds to the lowest energy state of the system (i.e., ground state). A quantum system can be mathematically represented by an Hamiltionian, i.e. a time-dependent operator that describes the system's total energy and whose eigenvalues correspond to the possible energy level the system can assume. 
The system Hamiltonian is
\begin{equation}
    H(t)= A(t)H_i+B(t)H_p
\end{equation}
The process starts by initializing the system in a simple, known ground state $H_i$; then, the system is evolved gradually by slowly changing its Hamiltionian from the initial one to the final one $H_p$, which represents the problem to be solved. Quantum tunneling allows this method to leap from one local minimum to another, while the quantum adiabatic theorem ensures that, if the perturbation acting on the system is sufficiently slow, then it will remain in a ground state \cite{Morita2008}.

\subsection{Neutral-atom quantum computers}

Neutral-atom quantum processors are a relatively new technology compared with more consolidated quantum annealers. They present unique characteristics, such as allowing high connectivity between qubits and a flexible, programmable architecture, which can be reasonably beneficial for practical implementations of complex optimization problems. These machines operate by rearranging Rydberg atoms, trapped in optical tweezers, and inducing dipole-dipole interaction via optical or microwave pulses \cite{Henriet2020}. This dipole-dipole interaction is better exploited using alkali atoms, such as Rubidium atoms.

In general, a Rydberg atom is an excited atom with one (or more) electrons with very high principal quantum number $n$. 
Alkali atoms have only a single electron in the outermost level, so exciting this electron will create a large electron dipole moment. 
If we place two excited Rydberg atoms close together (closer than the distance known as the Rydberg radius $R_b$, which is $\mathcal{O}(\mu \mathrm{m})$ for Rubidium), their electric dipole momenta start to interact with each other \cite{Gaetan:2009ele,Urban:2009hsg}. This interaction will cause a splitting of the atom's energy levels in such a way that it prevents the two atoms to be in the same excited state \cite{Henriet2020,Jaksch_2000,Lukin_2001}.

This is the fundamental mechanism exploited in neutral-atom machines and is known as Rydberg blockade. 
In an isolated Rydberg atom the energies of the ground state $\ket{g}$ and the excited state $\ket{r}$ differ by a value $\Omega$. When two atoms are put close enough the doubly excited state $\ket{rr}$ is not accessible anymore due to the dipole-dipole interaction: only the state $\ket{\psi_+} = 1/\sqrt{2}(\ket{gr}+\ket{rg})$ will be populated. If we consider the situation in which atoms are arranged in a lattice, an atom excited to the Rydberg level prohibits the excitation of any atom within a sphere of radius $R_b$.

Neutral-atom quantum processors are suitable for the implementation of the quantum CNOT gate. However, for the scenario where atoms are placed in a lattice, the interaction among Rydberg states can be translated into a Ising-type Hamiltonian
$H=\frac{1}{2} \sum_{i \neq j} \frac{C_3}{R_{ij}^3} \left(\sigma_i^+ \sigma_j^- + \sigma_i^- \sigma_j^+\right)$  where $\sigma^\pm$ are the Pauli matrices and $R_{ij}=|r_i-r_j|$ is the distance between atoms $i$ and~$j$ \cite{Henriet2020, Barredo:2015jzb}.


\section{Proposed method}

This section describes the proposed hybrid algorithm for clustering aggregation. The key idea is reducing the clustering aggregation problem to an equivalent graph partitioning problem, in particular the Maximum-Weight Independent Set (MWIS); this intuition was introduced in \cite{aggregation} and employed simulated annealing to solve the MWIS problem. The possible use case for a quantum machine relies on the fact that such a problem is NP-hard, and with classical computers, it can only be solved, in a reasonable amount of time, with approximate methods, such as the aforementioned simulated annealing; a quantum approach could potentially lead to quantum advantage. 

Additionally, it is possible to demonstrate that the MWIS is equivalent to a QUBO problem with constraints, a class of problems that can be tackled by D-Wave quantum annealers. Although this problem cannot be, at the moment, solved by Pasqal's Fresnel, this neutral-atom-based quantum processor can solve Maximum Independent Set (MIS) problems. Given that an MWIS is a MIS with additional constraints, Fresnel can be used to reduce the search space for finding an optimal clustering.

\subsection{Clustering aggregation as a graph problem}
\label{sec:aggregation_graph}

Given a dataset $D = \{d_{1},...,d_{n}\}$, we consider $m$ clusterings of $D$; each clustering $\mathcal{C}_{i}$ contains $k_{i}$ clusters, which are subsets of $D$.

It is possible to build an undirected graph $\mathcal{G} = (\mathcal{V}, \mathcal{E})$, where 
\begin{align}
  \label{eqn:union}
  \mathcal{V} &= \bigcup_{i = 1}^{m} \mathcal{C}_{i} = \{c_{11},...,c_{1k_{1}},...,c_{m1},...,c_{mk_{m}}\}\\
  \label{eqn:overlap}
  \mathcal{E} &= \{(c_{i}, c_{j}) \in \mathcal{V}^{2} \vert c_{i} \cap c_{j} \neq \emptyset \};
\end{align}
in other words, the vertices of the graph are the clusters from all the clusterings, and an edge is between two of them if and only if they overlap. According to Equation \ref{eqn:overlap}, a necessary condition for a clustering to be valid is that its clusters do not overlap; in the graph representation, this translates to the fact that vertices representing clusters of a valid clustering must form an independent set.

In this manner, the search space for the optimal aggregated clustering is reduced to the clusterings that form independent sets. To find the optimal one, it is necessary to choose a metric to maximize; for this algorithm, as proposed in \cite{aggregation}, the chosen metric is the sum of the total silhouette score of the clusters composing the clustering. 

In terms of graph representation, this is equal to associating a weight (the silhouette score) with each vertex, and the problem of finding the optimal clustering becomes a search for a Maximum-Weight Independent Set, an optimization problem on graphs that will be formally defined in Section \ref{sec:mwis}.

\subsection{QUBO formulation of the MWIS}

\label{sec:mwis}
A graph $\mathcal{G} = (\mathcal{V}, \mathcal{E})$ can be represented in matrix form via its associated adjacency matrix, which has the form $A = (a_{ij})_{m \times m}$, where
\begin{equation}
    a_{ij} = 
    \begin{cases}
        1 & \text{if } (i, j) \in \mathcal{E} \\
        0 & \text{otherwise},
    \end{cases}
    \label{eqn:adj_matrix}
\end{equation}
and $\vert \mathcal{V} \vert = m.$
Furthermore, a subset of the vertices $\mathcal{V}' \subseteq \mathcal{V}$ can be represented as a vector $\mathbf{x} \in \{0, 1\}^{m}$, where
\begin{equation}
    x_{i} = 
    \begin{cases}
        1 & \text{if } i  \in V' \\
        0 & \text{otherwise}.
    \end{cases}
\end{equation}

Calling $\mathbf{w} \in \mathbb{R}^{m}$ the vector of weights associated to each vertex, finding a MWIS in $\mathcal{G}$ is equivalent to finding a maximal vector $\mathbf{\overline{x}} \in \{0, 1\}^{m}$ satisfying
 \begin{equation}
     \mathbf{\overline{x}} = \argmax_{\mathbf{x}} \mathbf{w}^{T}\mathbf{x},
     \label{eqn:maximize}
\end{equation}
such that
\begin{equation}
    \mathbf{\overline{x}}^{T} A \mathbf{\overline{x}} = 0.
    \label{eqn:mis}
\end{equation}
Equation \ref{eqn:maximize} expresses the constraint that the chosen vertices must be those that maximize the sum of their associated weights. We also observe that the maximal vectors that solve Equation \ref{eqn:mis} are solutions to the 
MIS problem, which is an NP-hard optimization problem on graphs.

Because the graph is undirected, the matrix $A$ can be considered an upper triangular matrix, and Equation \ref{eqn:mis} expresses therefore a QUBO (Quadratic Unconstrained Binary Optimization) problem. Equation \ref{eqn:maximize} is an additional constraint for the QUBO problem.

Being able to reduce the clustering aggregation problem to MWIS, which in turn can be reduced to a QUBO with constraints, is essential for submitting the problem to the quantum platforms we are considering. While D-Wave's Advantage QPU can solve QUBO problems with constraints, Pasqal's Fresnel can currently only work in analog mode and solve MIS problems. Thus, when testing the algorithm, Fresnel can only find solutions that satisfy Equation \ref{eqn:mis}.

\subsection{Benchmarking of quantum computers}

Measuring the performance of quantum computers is not only an essential step in assessing how a single quantum computer compares to other quantum computers, but is also an important indicator of the extent to which these machines are capable of tackling practical problems, also referred to as \emph{quantum utility} \cite{Proctor2024}. 

Benchmarking of quantum systems, a term used to indicate the systematic process of evaluation and measurement of performances, presents numerous challenges, particularly when comparing different quantum technologies such as gate-based, annealing and neutral-atoms processors. These systems operate based on different physical principles, making it difficult to establish a universal metric for comparison. Another factor affecting the complexity of performing a benchmark of quantum processors is that certain technologies are suited to work only with a restricted set of tasks; designing a single algorithm to serve as a benchmark for a generic quantum computer would not consider the differences in the various approaches. For example, quantum annealers, which rely on adiabatic evolution to solve optimization problems, excel at specific tasks, but are hindered by issues such as noise, qubit connectivity, and problem embedding, which limit their scalability. Similarly, neutral-atom quantum processors, which can, theoretically, work in both digital (gate-based) and analog mode \cite{Wintersperger2023}, offer highly scalable architectures by manipulating individual atoms through optical tweezers, which mitigates the issue of problem embedding, but have still to face noise, decoherence and fidelity during quantum operations \cite{Henriet2020}. Gate-based systems are, on the other hand, more versatile and capable of working with a broader set of algorithms but suffer from qubit decoherence and operational errors, which reduce gate fidelity in multi-qubit operations \cite{Moll2018}.


Solving clustering aggregation may be a valid approach to the benchmarking of neutral-atoms and annealing quantum processors, two technologies for which benchmarking research is lacking, given that the majority of research in the field is focused on gate-based quantum computers \cite{Barbaresco2024}.

Currently, Pasqal's neutral-atom-based Fresnel is only able to work in analog mode. It is only possible to control all qubits at the same time, but not individually; the only classes of optimization problems addressable right now by this machine are those that can be reduced to MIS, while MWIS problems are not easily solvable. For this reason, the information of the silhouette cannot be included in the problem embedding.

D-Wave's quantum annealer, Advantage QPU, on the other hand, offers a framework to add constraints to the QUBO sent as input to the solver, making it possible to fully perform the intended algorithm and solve the MWIS problem.

\subsection{Description of the algorithm}
\begin{figure}[ht!]
    \centering
    \includegraphics[width=0.8\linewidth]{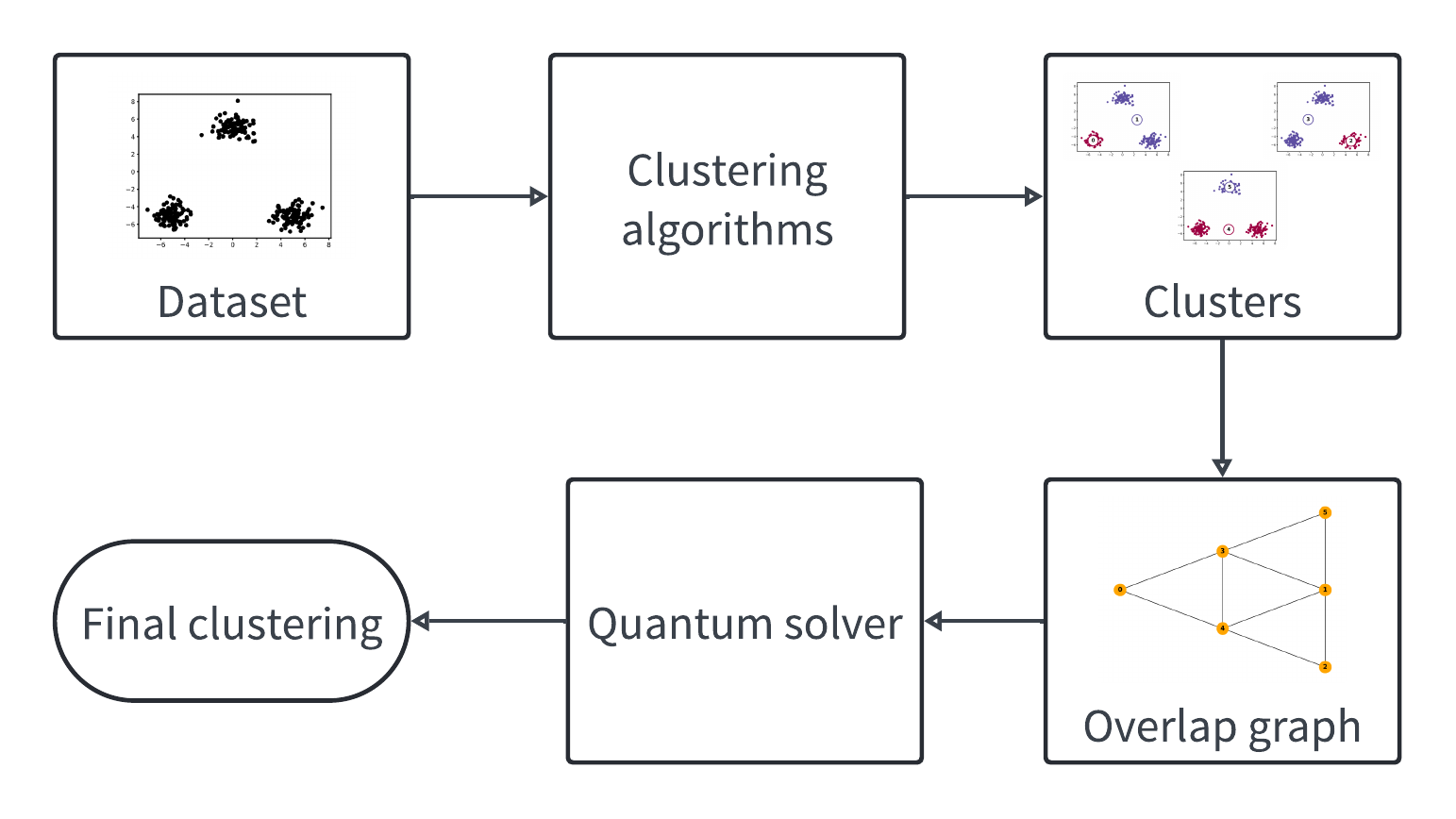}
    \caption{Steps of the proposed algorithm.}
    \label{fig:proposed_algorithm}
\end{figure}

The steps of the proposed method are listed below and are visualized in Figure \ref{fig:proposed_algorithm}:
\begin{description}
  \item[Step 1] Multiple clustering algorithms, either distinct algorithms such as DBSCAN and Spectral Clustering, or the same algorithm with different values of parameters (an example of algorithms with some of their parameters, from the \texttt{sklearn.cluster} implementation \cite{ScikitCluster2024}, are presented in Table \ref{tab:hyperparameters}), are run on the dataset; clusters produced by the various algorithms are then given unique labels and silhouette score is computed for each of them. In Table \ref{tab:hyperparameters} the parameters tuned in the practical implementation are indicated in bold font. K-means and Spectral Clustering require specification of the number of clusters into which the dataset should be partitioned, whereas DBSCAN's parameter, called epsilon, is a distance value used to select points to add to the clusters.
  \item[Step 2] Overlap graph is built according to Section \ref{sec:aggregation_graph}, with vertices of the graph weighted according to the silhouette score of the cluster they represent, and edges having associated weight equal to 1.
  \item[Step 3] Associated adjacency matrix is built from the overlap graph, in such a way it is triangular and all elements on the diagonal having the same value (possible because the graph is undirected).
  \item[Step 4] The QUBO matrix, built in the previous step from the overlap graph and representing the MWIS problem, is solved using a quantum machine.
\end{description}

\begin{table}[ht!]
    \centering
    \resizebox{\columnwidth}{!}{%
    \begin{tabular}{||l|l||}
        \hline 
        \textbf{Algorithm} & \textbf{Parameters} \\
        \hline 
        K-means & \textbf{number of clusters}, number of iterations, tolerance, ... \\
        \hline 
        DBSCAN & \textbf{epsilon}, minimum number of samples, leaf size, ... \\ 
        \hline 
        Spectral Clustering & \textbf{number of clusters}, gamma, affinity, ... \\
        \hline
    \end{tabular}
    }
    \caption{Algorithms used in this work with some of their corresponding parameters, in the \texttt{sklearn.cluster} implementation \cite{ScikitCluster2024}.  }
    \label{tab:hyperparameters}
\end{table}

Step $4$ of the algorithm is fundamentally dependent on the technology of the quantum machine chosen to perform the algorithm. While this problem could be solved using gate-based quantum computers, as well as classical emulators for this machines, for this work, two different approaches were chosen: a neutral-atoms quantum computer in analog mode, developed by Pasqal, and a D-Wave quantum annealer. 

\subsection{Example}\label{sec:example}

\begin{figure}[ht!]
  \centering 
  \includegraphics[width=0.9\linewidth]{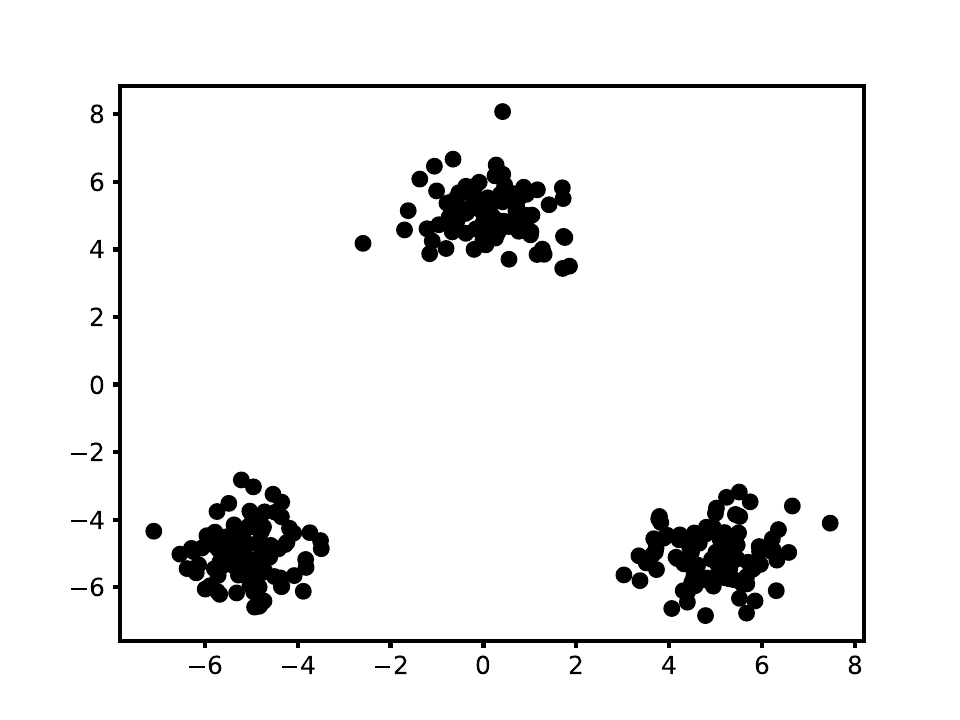}
  \caption{Example dataset, generated with function \texttt{make\_blobs} from module \texttt{sklearn.dataset}, containing 300 points divided in three blobs.}
  \label{fig:example_dataset}
\end{figure}

We present here a simple example of the algorithm, tested on a small dataset, shown in Figure \ref{fig:example_dataset}. The dataset was generated using the \texttt{make\_blobs} function from the \texttt{sklearn.dataset} module \cite{Scikit2024} and contains 300 points divided into three blobs, positioned at equal distances from one another.

\begin{figure*}[ht!]
  \centering
  \subfigure[Clustering \#1]{
    \includegraphics[width=0.3\textwidth]{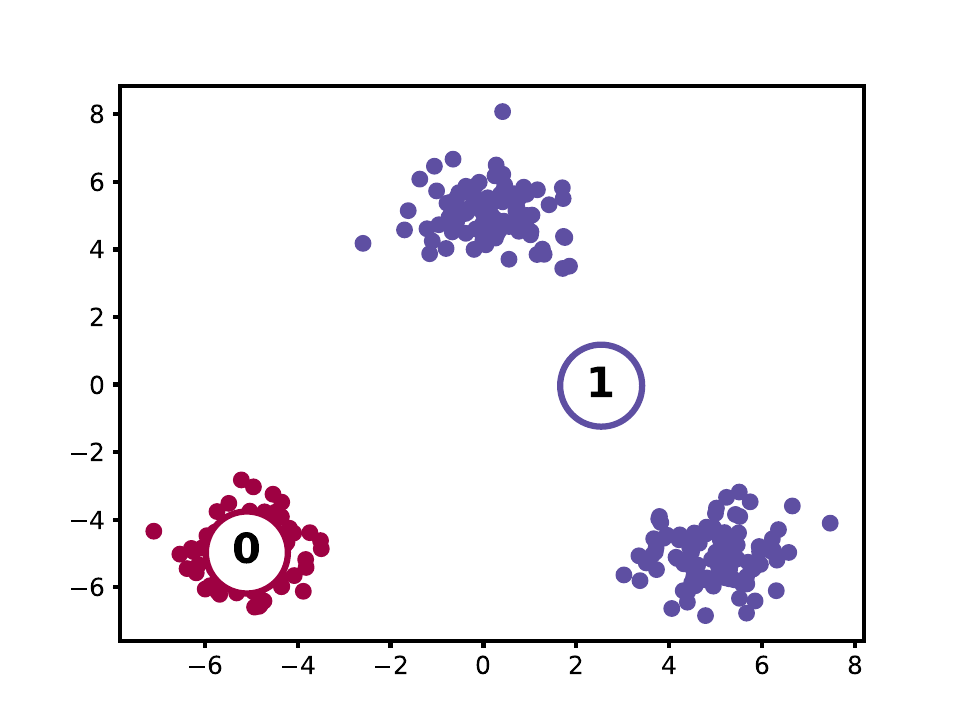}
  }
  \subfigure[Clustering \#2]{
    \includegraphics[width=0.3\textwidth]{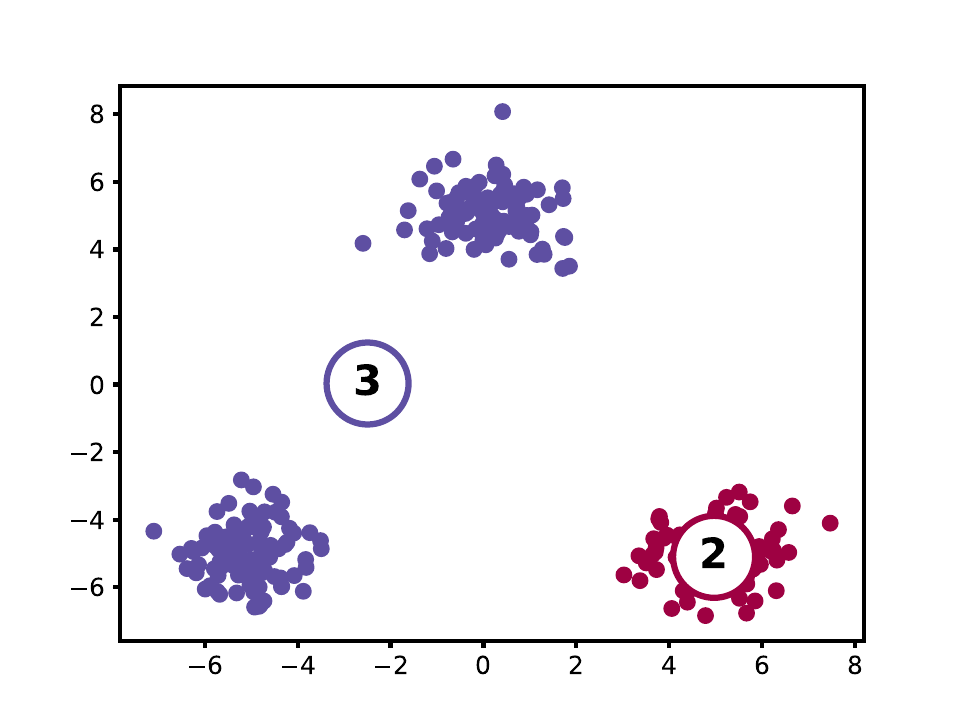}
  }
  \subfigure[Clustering \#3]{
    \includegraphics[width=0.3\textwidth]{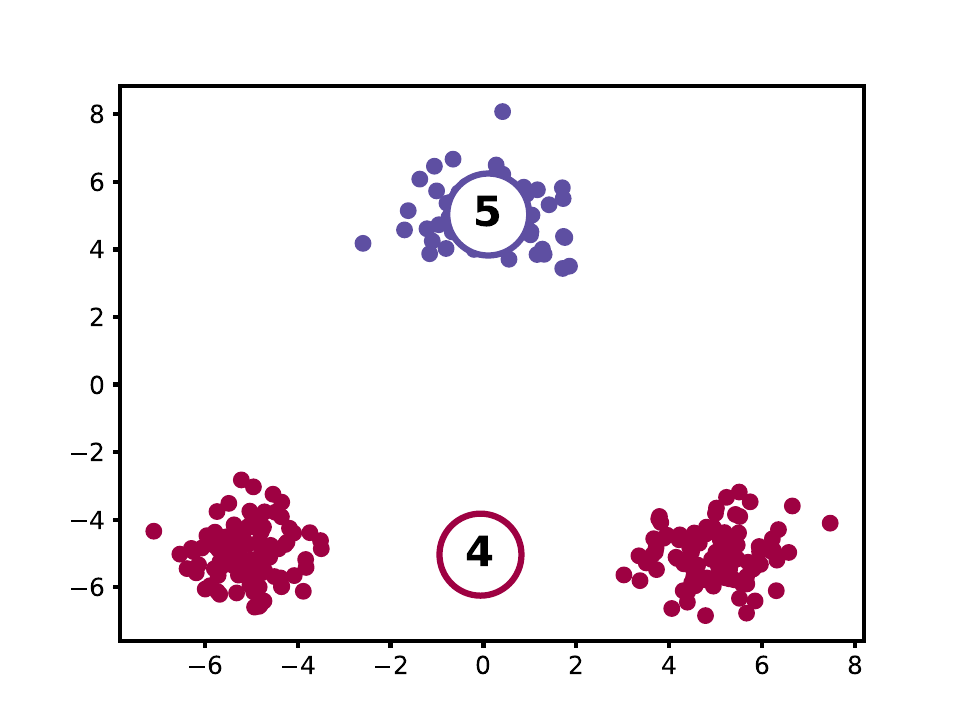}
  }
  \caption{Three clusterings of the original dataset, all of which obtained via K-Means with number of clusters equal to 2 and seed chosen in order to obtain different results. (a) contains clusters with labels 0 and 1; (b) contains clusters with labels 2 and 3; (c) contains clusters with labels 4 and 5.}
  \label{fig:example_clusterings}
\end{figure*}

The first step is running on the dataset multiple clustering algorithms, which can be different algorithms (such as DBSCAN or Spectral Clustering), or the same algorithm with different hyperparameters. In this case, three instances of K-Means were run on the dataset (Figure \ref{fig:example_clusterings}), all of them with number of clusters set to 2 but having different seed values, chosen in order to obtain three distinct results. In this toy example, the seed is playing the role of a generic hyperparameter. Each clustering produces two clusters, which are uniquely labeled across the various clusterings; silhouette scores are then computed for each point, and each cluster is assigned the sum of the silhouettes of the point it contains (Table \ref{tab:silhouette_scores}).

\begin{table}[ht!]
  \centering
  \begin{tabular}{||l|l|l||}
    \hline
    Cluster & Silhouette (avg) & Silhouette (sum) \\
    \hline 
    0 & 0.875 & 87.529 \\
    1 & 0.402 & 80.328 \\
    2 & 0.867 & 86.724 \\
    3 & 0.402 & 80.381 \\
    4 & 0.489 & 97.723 \\
    5 & 0.878 & 87.761 \\
    \hline
  \end{tabular}
  \caption{Silhouette scores for the example clusterings.}
  \label{tab:silhouette_scores}
\end{table}

\begin{figure}[ht!]
  \centering 
  \includegraphics[width=0.5\linewidth]{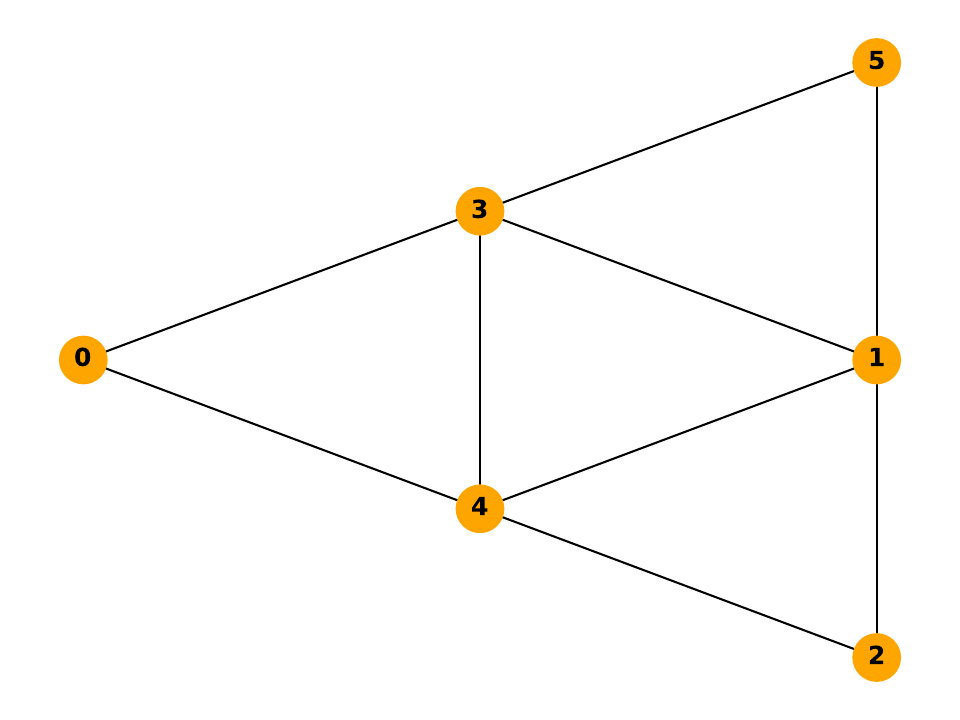}
  \caption{Adjacency graph for the example dataset.}
  \label{fig:adj_example}
\end{figure}

An overlap graph is then built according to the points shared between clusters, as shown in Figure \ref{fig:adj_example}. Its associated adjacency matrix representation has the form
\begin{equation}
  A = 
  \begin{bmatrix}
    1 & 0 & 0 & 1 & 1 & 0 \\
    0 & 1 & 1 & 1 & 1 & 1 \\
    0 & 0 & 1 & 0 & 1 & 0 \\
    0 & 0 & 0 & 1 & 1 & 1 \\
    0 & 0 & 0 & 0 & 1 & 0 \\
    0 & 0 & 0 & 0 & 0 & 1 \\
  \end{bmatrix}.
\end{equation}

The only independent set that satisfies both Equation \ref{eqn:union} and Equation \ref{eqn:overlap} is the vector $\mathbf{x} = \begin{bmatrix} 1 & 0 & 1 & 0 & 0 & 1 \end{bmatrix}$, representing the clustering which contains clusters with labels 0, 2 and 5. In this particular case, being this the only solution, it is not necessary to check which solution maximizes the silhouette sum.

\section{Experiments and results}
\subsection{Dataset}
\label{sec:dataset}
\begin{figure}[ht!]
  \centering 
  \includegraphics[width=0.75\linewidth]{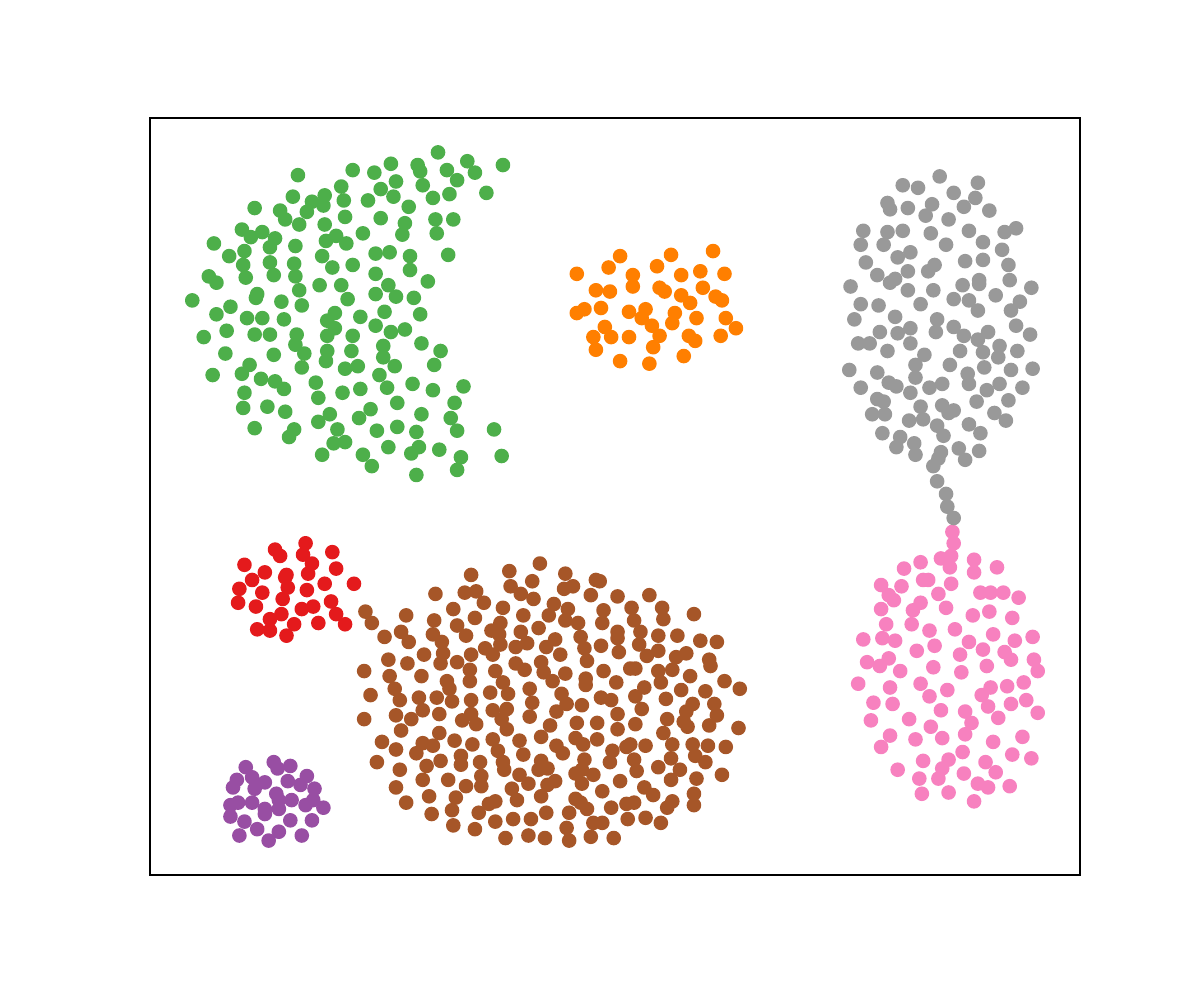}
  \caption{Plot of the dataset used to test the hybrid clustering aggregation algorithm, made up of 788 points. The different colors indicate the 7 clusters of the clustering considered optimal.}
  \label{fig:dataset}
\end{figure}

The dataset used to test the hybrid clustering aggregation is shown in Figure \ref{fig:dataset} and was first introduced by Gionis, Mannila and Tsaparas to test classical clustering aggregation algorithms \cite{Gionis2007}; it was then used by Li and Latecki to test a classical clustering aggregation algorithm that uses simulated annealing \cite{aggregation}.

\begin{figure*}[ht!]
    \centering
    \subfigure[]{
        \centering
        \includegraphics[width=0.4\textwidth]{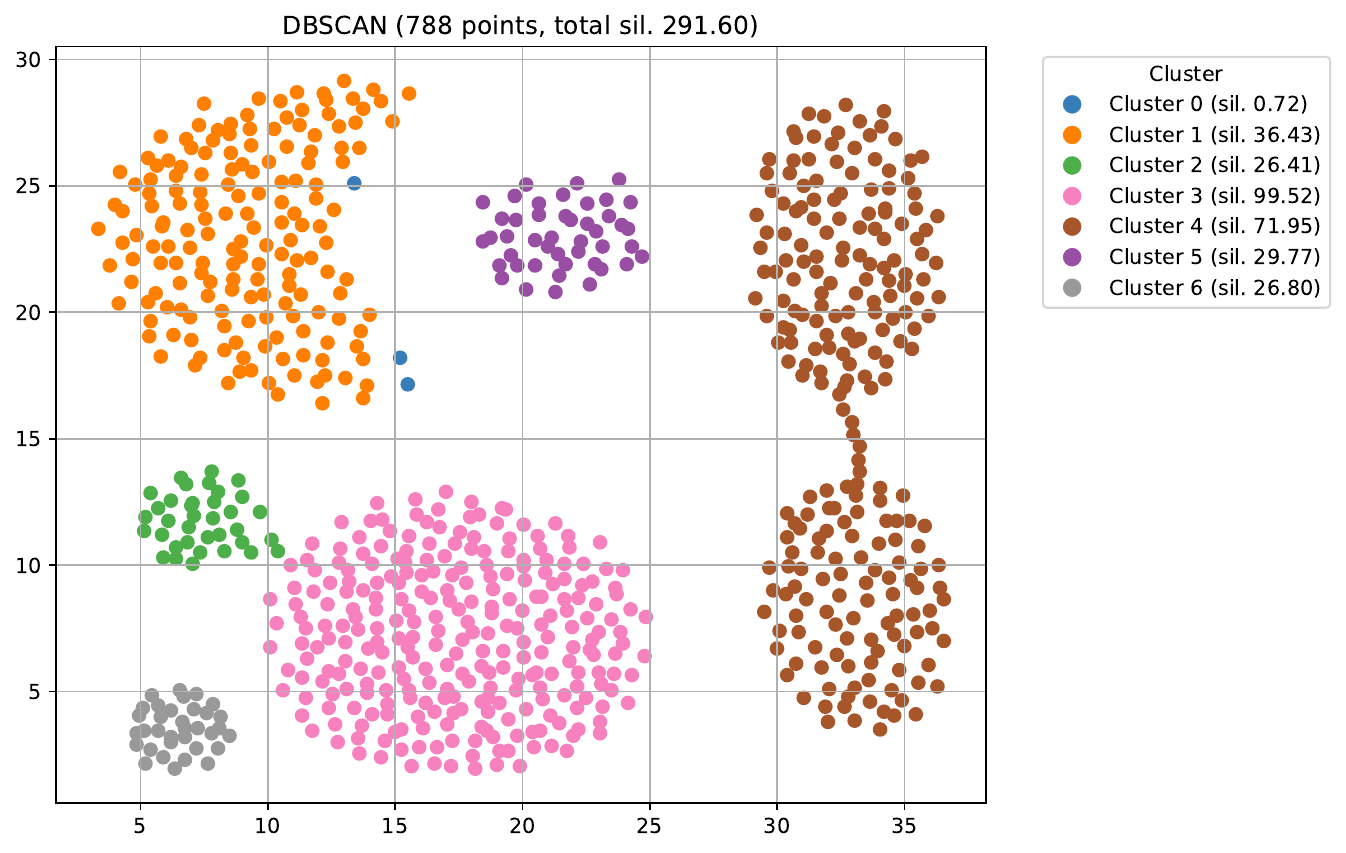}
    }
    \subfigure[]{
        \centering
        \includegraphics[width=0.4\textwidth]{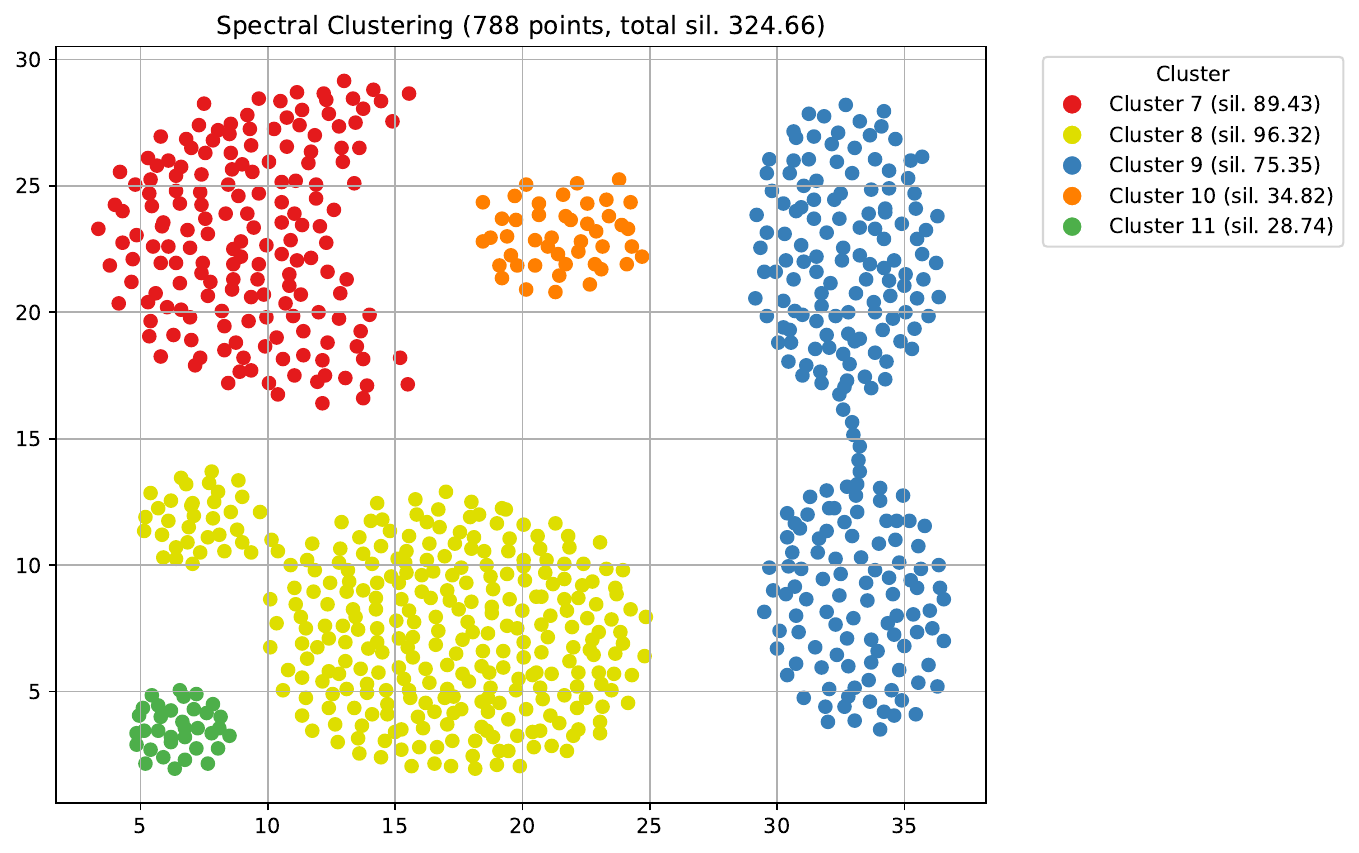}
    }
    \caption{Results of the clustering algorithms run on the input dataset, whose clusters will be selected by the clustering aggregation algorithm. Clustering in (a) was obtained through DBSCAN, with epsilon equal to 1.2, and contains a total of 7 clusters labeled from 0 to 6; clustering in (b) was obtained through Spectral Clustering, with number of clusters equal to 5, and contains 5 clusters labeled from 7 to 11.}
    \label{fig:algorithms}
\end{figure*}

The classical algorithms chosen for the first step of the proposed method are DBSCAN, with parameter epsilon equal to 1.2, and Spectral Clustering, with parameter number of clusters equal to 5, which produced clusterings with 7 and 5 clusters, respectively, for a total of 12 clusters  (we label these clusters with an increasing number, from $0$ to $11$, as shown in Figure \ref{fig:algorithms}). Since the total number of clusters corresponds to the size of the overlap graph and, subsequently, the number of required qubits, the amount of clustering algorithms and their relative parameters were chosen with the aim of preventing this number to grow too large, because of the number of qubits available on Fresnel. While Fresnel can work with up to 100 qubits \cite{Fresnel2024}, at the time of writing they can only be placed on a fixed layout. Due to intrinsic limits of Rydberg atoms (namely, that they have to be at a distance larger than the Rydberg radius to prevent unnecessary entanglements, but cannot be closer than $5 \mu m$), it was simpler to consider a smaller, yet significant, instance of the problem, which required 12 total qubits.

\subsection{Experiments setup}

A first step of validation was performed on a smaller dataset, in order to determine the feasibility of the algorithm on Fresnel. After that, the algorithm was tested on the dataset described in Section \ref{sec:dataset}, in order to compare the performances of Fresnel and Advantage QPU.

The clusterings returned by the quantum platforms are in the form of 12-digit bitstrings: if, for example, the found solution is completely aligned with the one found by DBSCAN, the bitstring will be $111111100000$. On the contrary, if the found solution contains the same clusters obtained by running Spectral Clustering, the bitstring will be $000000011111$. A generic solution, for example, with bitstring $001010011001$, corresponds to a situation where clusters labeled $2$ and $4$ are taken from DBSCAN, whereas clusters labeled $7$, $8$ and $11$ come from Spectral Clustering. 

Running the experiments on Fresnel requires to define a qubit configuration that encodes the problem in the register, and specify a sequence of pulses to act on the system of qubits in order to make it reach its ground state. Both the register configuration and the sequence are the same for the experiment on QuTiP and tensor networks emulators and the one on real hardware.

Due to the impossibility, at the moment of the experiments, of controlling each qubit independently, currently there are no means to add the silhouette score information to weigh vertices of the graph; the algorithm can therefore only solve a MIS problem (and not a MWIS, which is equivalent to clustering aggregation).

The sequence used was built with the Pulser framework \cite{Silverio2021}, and is made up of a single pulse; the Rabi frequency function $\Omega(t)$ is obtained with the Pulser object \texttt{InterpolatedWaveform}, constructed with values \texttt{T, [1e-9, Omega, 1e-9]}, while the detuning function $\delta(t)$ with the same object, but with parameters \texttt{T, [delta\_0, 0, -delta\_0]}. The values of these parameters are displayed in Table \ref{tab:sequence_parameters}.


\begin{table}[ht!]
    \centering
    \begin{tabular}{||l|l||}
        \hline
         \textbf{Parameter} & \textbf{Value}  \\
         \hline 
         \texttt{Omega} & 0.9552 \\ 
         \hline 
         \texttt{delta\_0} & -3 \\
         \hline 
         \texttt{T} & 3873 \\
         \hline
    \end{tabular}
    \caption{Parameters used to generate the Rabi frequency and detuning functions of the pulse contained in the sequence used for the experiments on Fresnel.}
    \label{tab:sequence_parameters}
\end{table}

Values of the parameters used to generate the sequence, namely \texttt{Omega}, \texttt{delta\_0} and \texttt{T}, were obtained through Bayesian optimization. In particular, the function used was \texttt{gp\_minimize}, contained in the package \texttt{skopt} \cite{Bayesian2024}, which found the parameters of the sequence that yielded clusterings with the highest total silhouette score, after a total of 10 iterations.

Experiments on D-Wave's Advantage QPU did not require the same manual register preparation of Fresnel, given that the Advantage QPU has a fixed topology; on the other hand, the embedding of the graph onto the machine's topology is not a trivial taks and represents a huge bottleneck of the process. The possibility of adding constraints to the problem makes it possible to test the algorithm both without providing the information of silhouettes, in order to compare it to the experiments on Fresnel with equal conditions, and with the silhouette values as well.
Both experiments were performed 1000 times, and results gathered in the form of bitstrings of length 12.

\subsection{Metrics chosen to compare results}

\subsubsection{Time} 
\begin{figure}[ht!]
    \centering
    \includegraphics[width=0.7\linewidth]{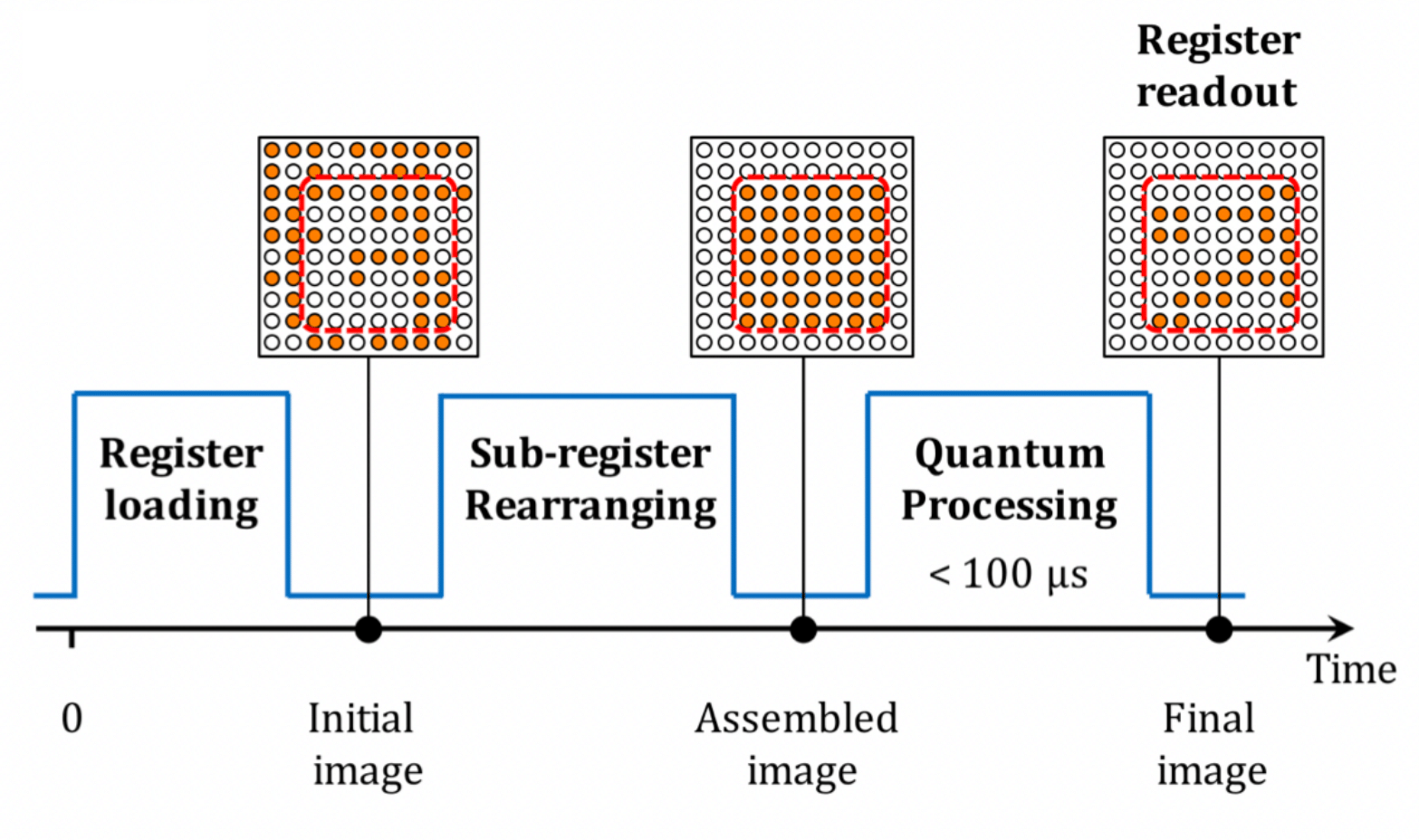}
    \caption{Steps of the QPU processing; at the moment, only quantum processing time has been disclosed, with a maximum value of $100 \mu s$.}
    \label{fig:fresnel_qpu_processing}
\end{figure}

An important detail to consider when comparing the efficiency of machines is the total time necessary to execute the algorithm. Due to the inherent differences of the quantum platforms used to perform the experiments, defining how to measure the execution time is not obvious, and requires a more thorough discussion. The processing cycle for Pasqal's Fresnel, shown in Figure \ref{fig:fresnel_qpu_processing}, is made up of the following steps: register loading, sub-register rearranging, and quantum processing. Unfortunately, at the moment the only disclosed information regarding timing is that the quantum processing step in a processing cycle takes at most $100 \mu s$. 
For the emulated version of Fresnel on QuTip, the execution time was obtained by calling Python's function \texttt{time.time()}, which returns the current UTC time, before and after the execution of the quantum simulation; execution time was then computed by taking the difference between the two timestamps. For the emulated version of Fresnel on tensor networks, which runs in  Pasqal's cloud infrastructure, the time information is provided in the results.

\subsubsection{Number of returned clusters}
The metric chosen to evaluate the quality of solutions is the probability of returning a solution with a certain number of clusters; in particular, a good algorithm should return the correct number of clusters (i.e. the size of the MWIS) with the highest probability.

\subsection{Validation}

\begin{figure}
    \centering
    \includegraphics[width=0.6\linewidth]{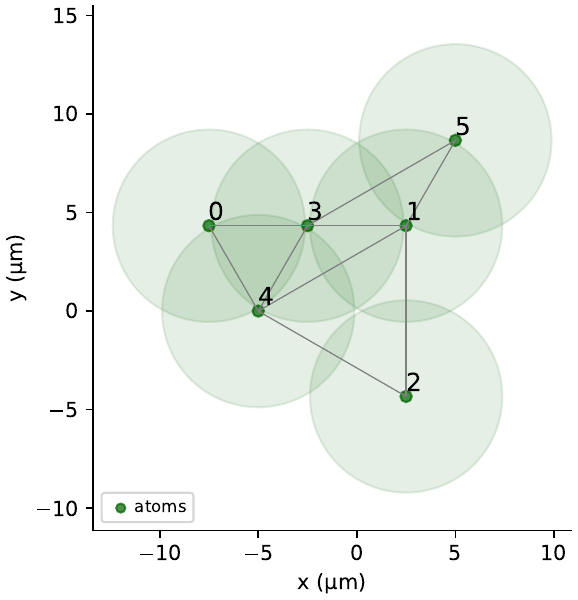}
    \caption{Plot of the qubits configuretion in the register of Fresnel for the first validation experiment.}
    \label{fig:register_simple}
\end{figure}

In order to validate the feasibility of the algorithm, a first experiment has been run on the simple example introduced in Section \ref{sec:example}. A first experiment was run on a QuTiP emulation of Fresnel; then, it was tested on real hardware Fresnel. The register configuration for the experiments, both on the emulator and the real hardware, is shown in Figure \ref{fig:register_simple}, and is topologically equivalent to the adjacency graph shown in Figure \ref{fig:adj_example}. The experiment on QuTiP emulator has an execution time of $1.3\cdot10^{5} \mu s$, while that on Fresnel, as discussed in the previous section, is estimated to be at most in the order of $10^{5} \mu s$.

\begin{figure}[ht!]
    \centering
    \subfigure[QuTiP emulated Fresnel.]{
        \includegraphics[width=0.9\linewidth]{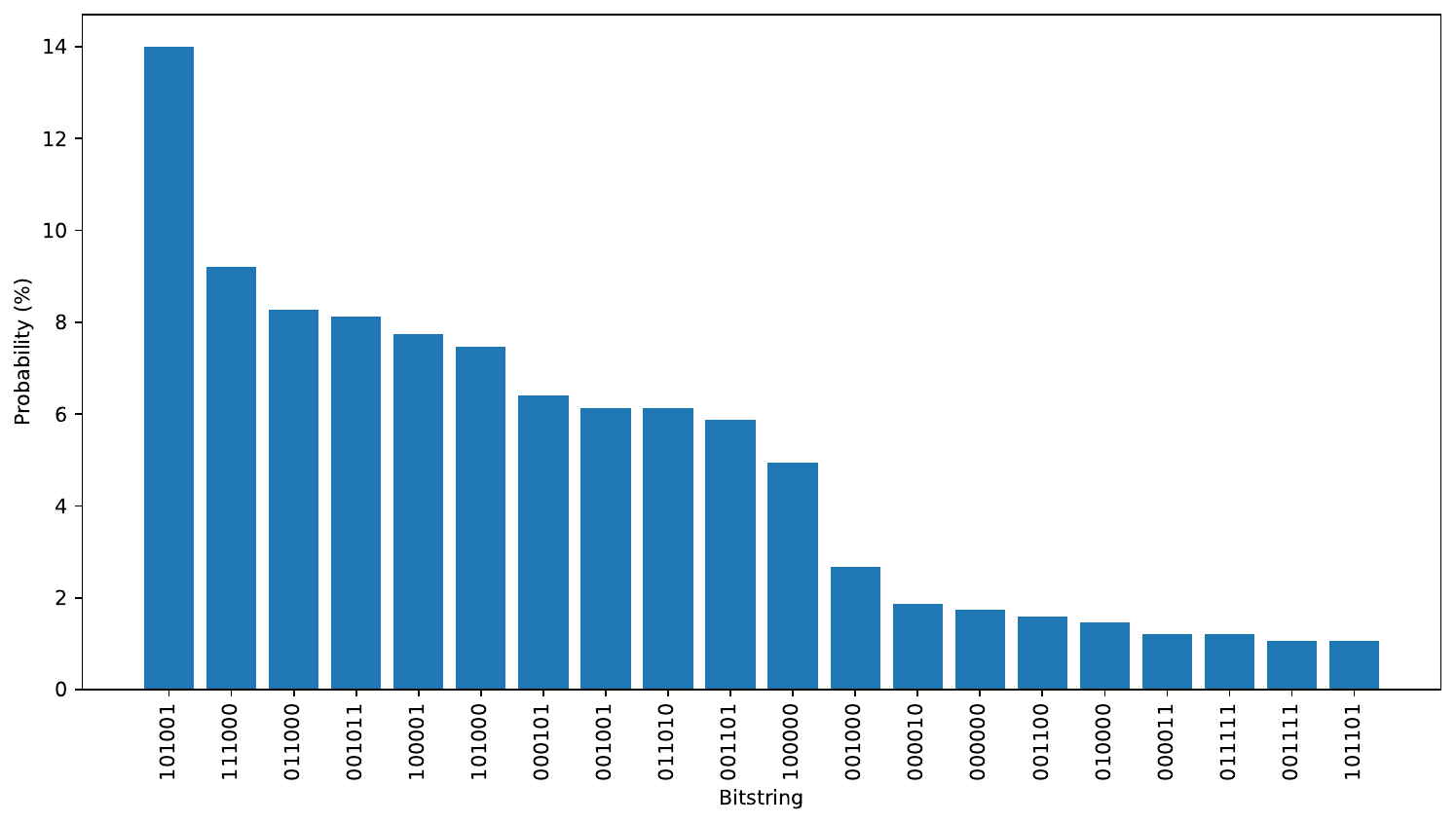}
    }

    \subfigure[Real hardware Fresnel.]{
        \includegraphics[width=0.9\linewidth]{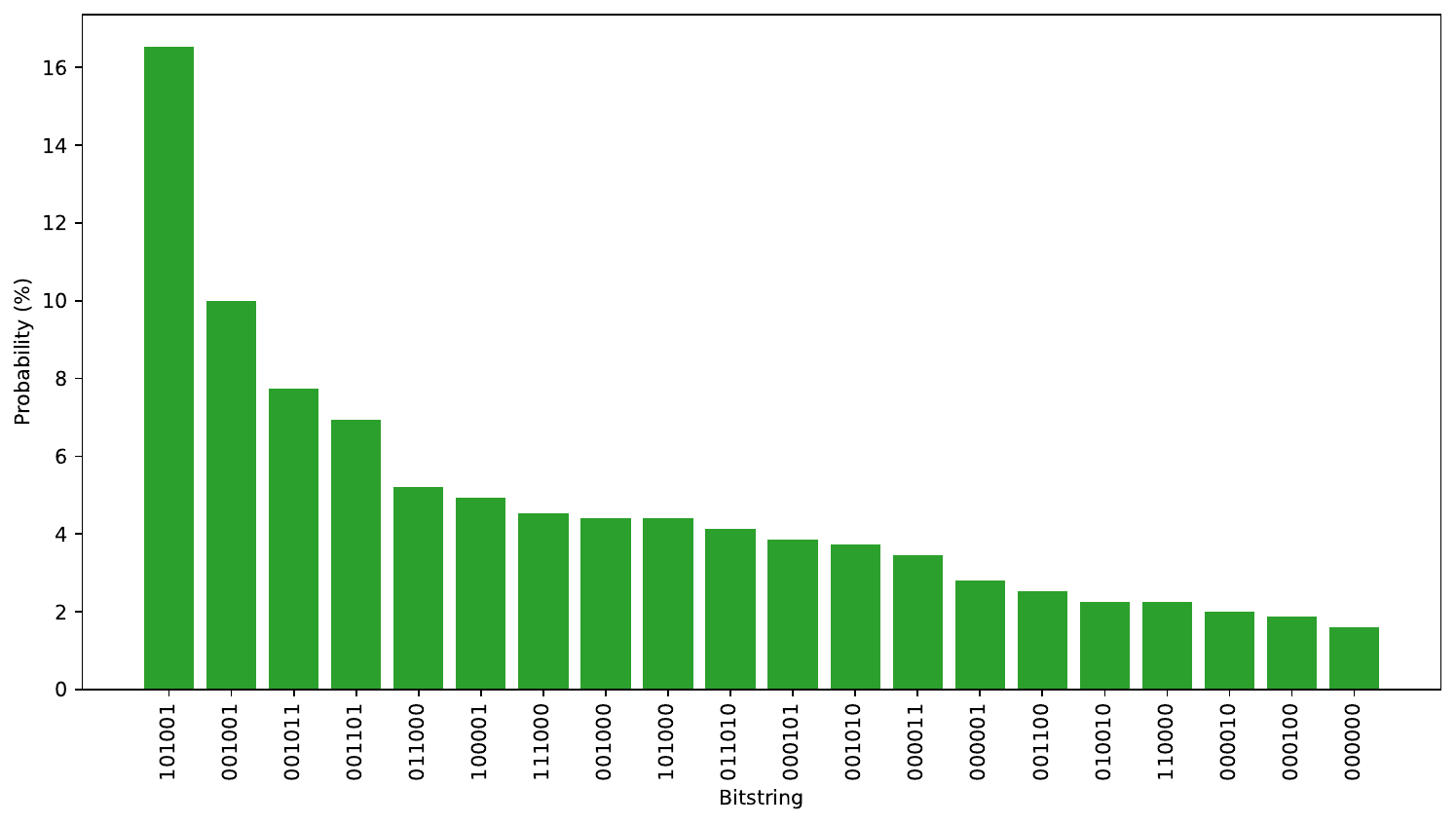}
    }
    \caption{Probability (in percentage) of reading a certain bistring after measuring the experiments on Fresnel (emulation on QuTiP and real hardware).}
    \label{fig:occurrences_simple}
\end{figure}

\begin{figure}
    \centering
    \includegraphics[width=0.8\linewidth]{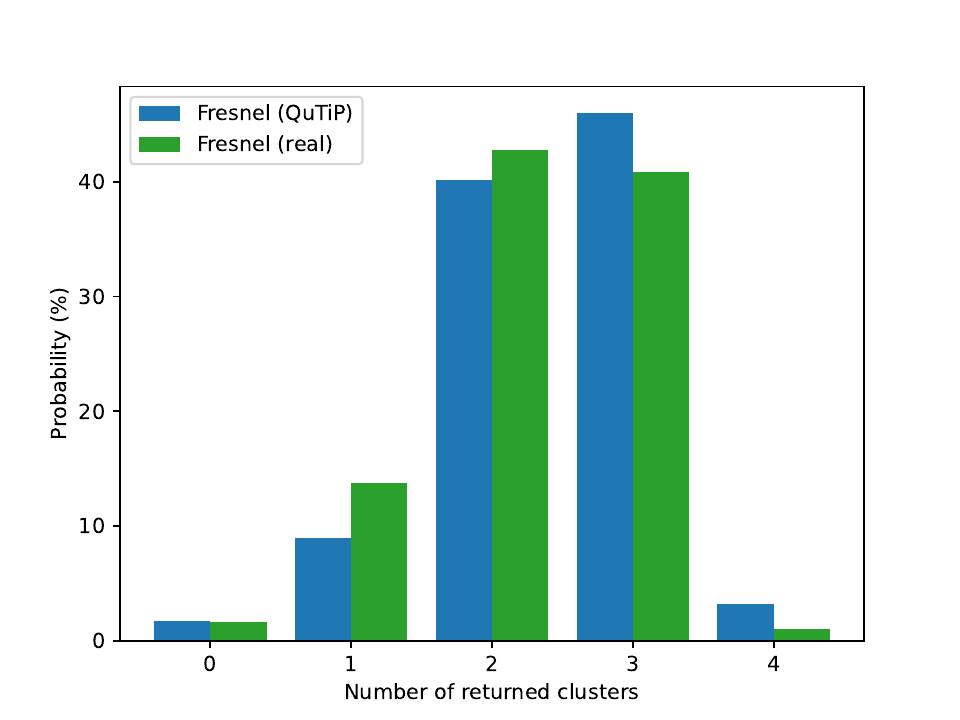}
    \caption{Percentage of bitstrings returning a certain number of clusters of the experiments onf Fresnel (emulation on QuTiP and real hardware).}
    \label{fig:num_clusters_simple}
\end{figure}

Figure \ref{fig:occurrences_simple} shows the most probable bitstrings returned in both experiments, while Figure \ref{fig:num_clusters_simple} shows the percentage of bistrings that return a certain number of clusters. In both cases, the most probable bitstring corresponds to the optimal solution of the problem (i.e. the only MIS). Furthermore, in both experiments arount 40\% of the bitstrings read returned a number of clusters equal to 3 (which correspons the the correct size of the MIS).

\subsection{Towards benchmarking}



After assessing the performance of Fresnel on a simple experiment, a second round of experiments has been run on a more complex dataset, which is the one presented in Section \ref{sec:dataset}. A total of four experiments were performed: one on real hardware Fresnel, one on tensor network emulator, and two on the Advantage QPU; the experiment were repeated 1000 times on each platform. The classical part of the algorithm, namely running the classical clustering algorithms on the input dataset, evaluating silhouettes, and building the overlap graph and associated adjacency matrix, was executed only once, and the results then sent to the different quantum platforms. 



\begin{figure}[ht!]
    \centering
    \includegraphics[width=0.8\linewidth]{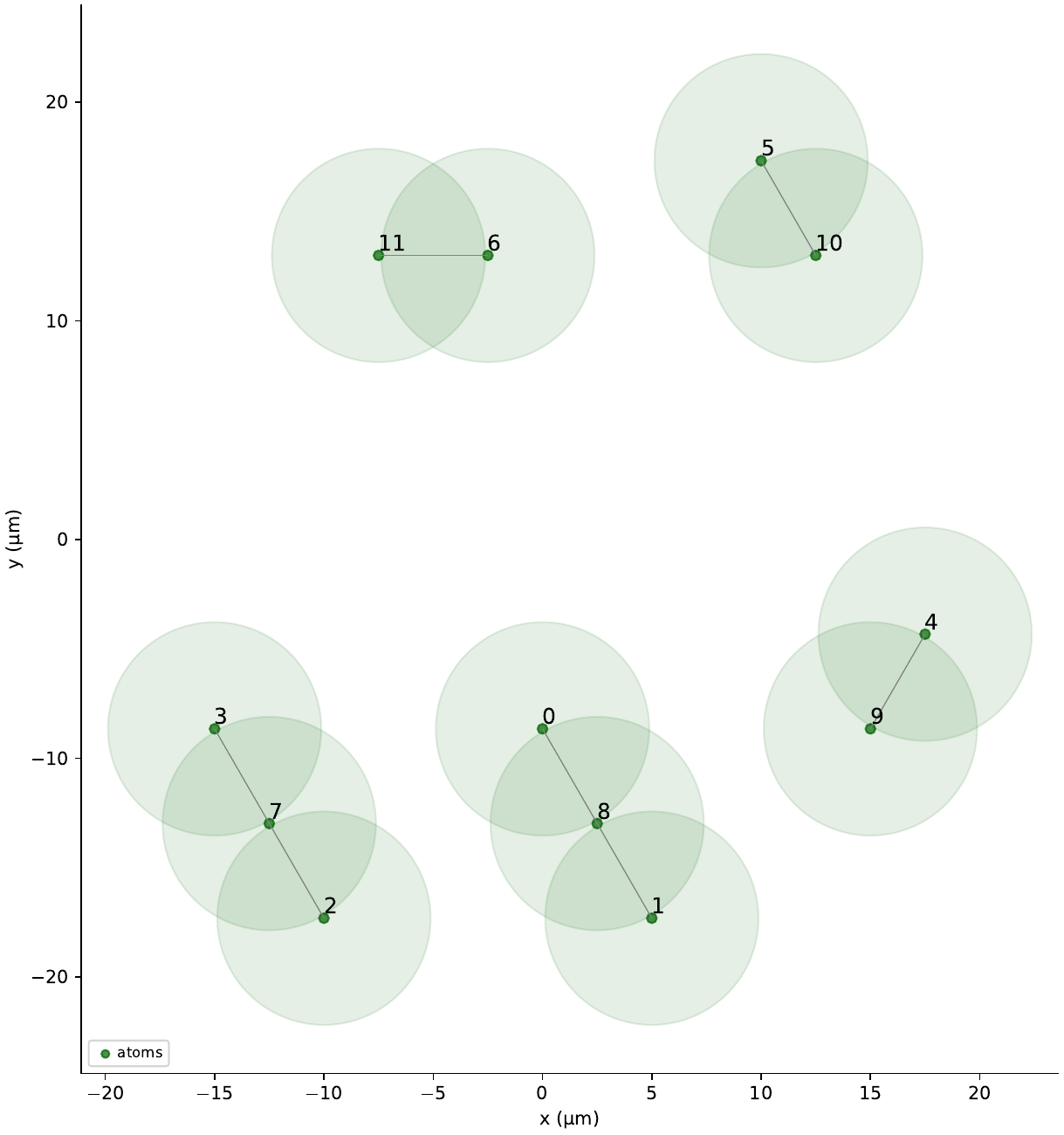}
    \caption{Plot of the qubits configuration in the register of Fresnel, with labels corresponding to clusters with the same label. This configuration was used both in the tensor network emulated Fresnel experiment, as well as the real hardware Fresnel experiment, and reflects the exact position of atoms. The green circles around qubits represent the Rydberg radius, while the arcs between qubits indicate that the qubits are entangled.}
    \label{fig:register}
\end{figure}

The register configuration for Fresnel, shown in Figure \ref{fig:register}, was built in such a way that qubits corresponding to overlapping clusters are at a Rydberg radius distance from one another, therefore entangled, recreating in the overlap graph obtained from the previous step of the algorithm.

A comparison of the execution times on the different platforms is shown in Table \ref{tab:time_comparisons}. It is possible to observe how the times of experiments run on real quantum hardware have comparable orders of magnitude (with the experiment on Advantage QPU without silhouette constraint being slightly faster, in the order of $10^{4} \mu s$), while that of the experiment on the tensor network emulator for Fresnel is higher (around $10^{7} \mu s$).

\begin{figure}[ht!]
    \centering
    \subfigure[QuTiP emulated Fresnel.]{
        \includegraphics[width=0.9\linewidth]{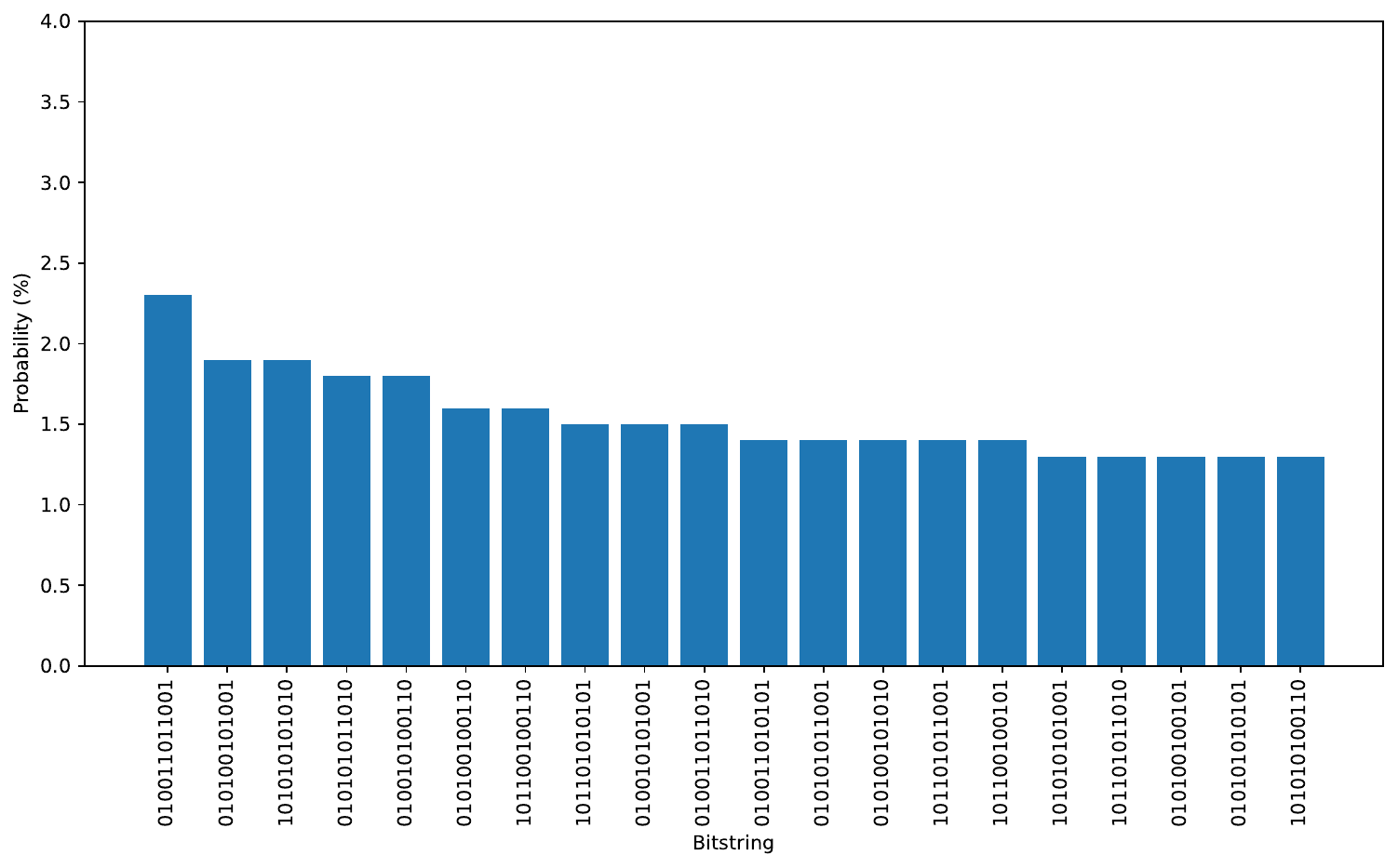}
    }

    \subfigure[Real hardware Fresnel.]{
        \includegraphics[width=0.9\linewidth]{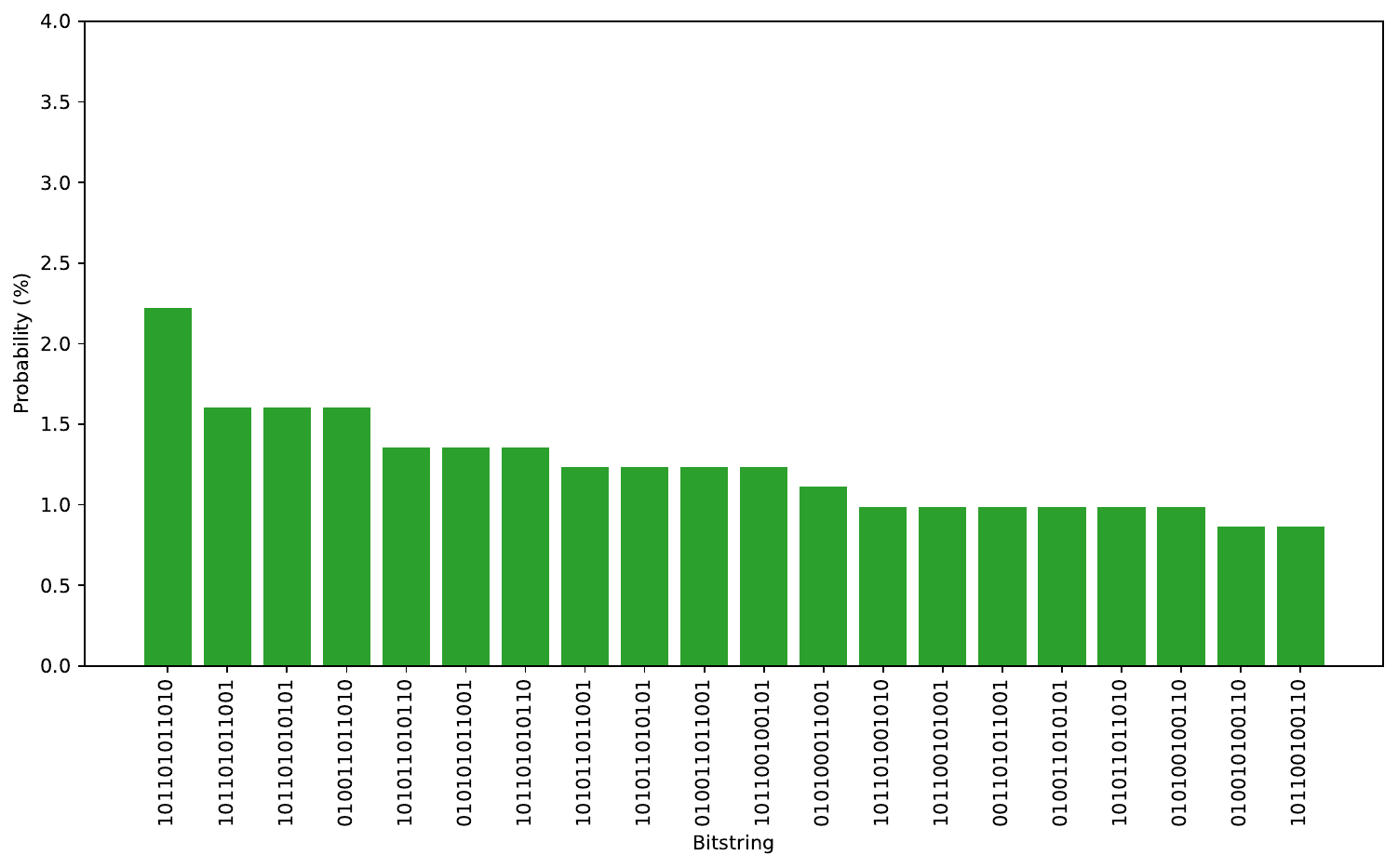}
    }
    \caption{Probability (in percentage) of reading a certain bistring after measuring the experiments on QuTiP emulated and real hardware Fresnel. (a) shows the 20 most likely bitstrings from the experiment on the emulator, while (b) the 20 most likely bitstrings from the real hardware experiment.}
    \label{fig:occurrences_pasqal}
\end{figure}

\begin{figure}[ht!]
    \centering
    \subfigure[Advantage QPU (without silhouette).]{
        \includegraphics[width=0.9\linewidth]{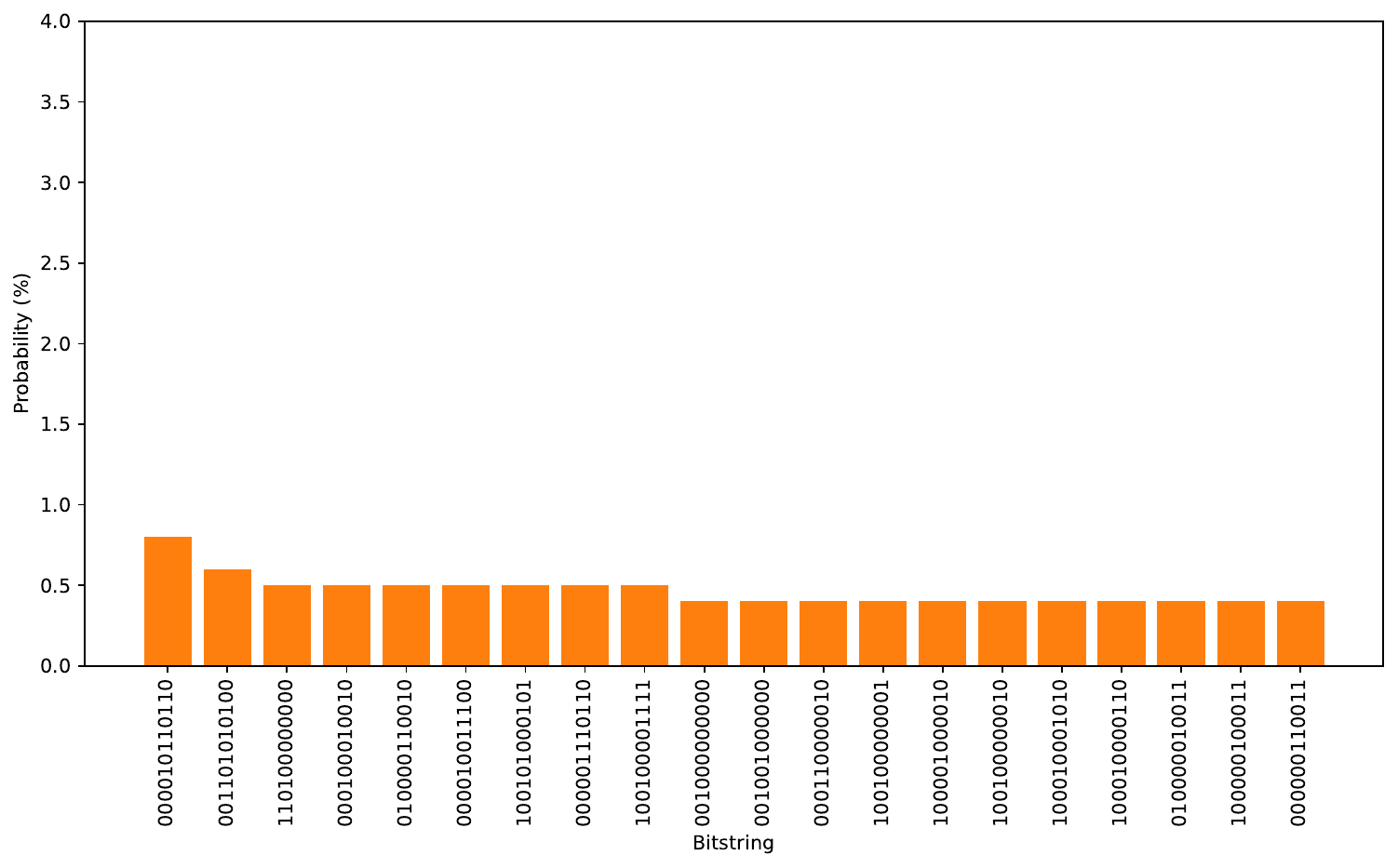}
    }

    \subfigure[Advantage QPU (with silhouette).]{
        \includegraphics[width=0.9\linewidth]{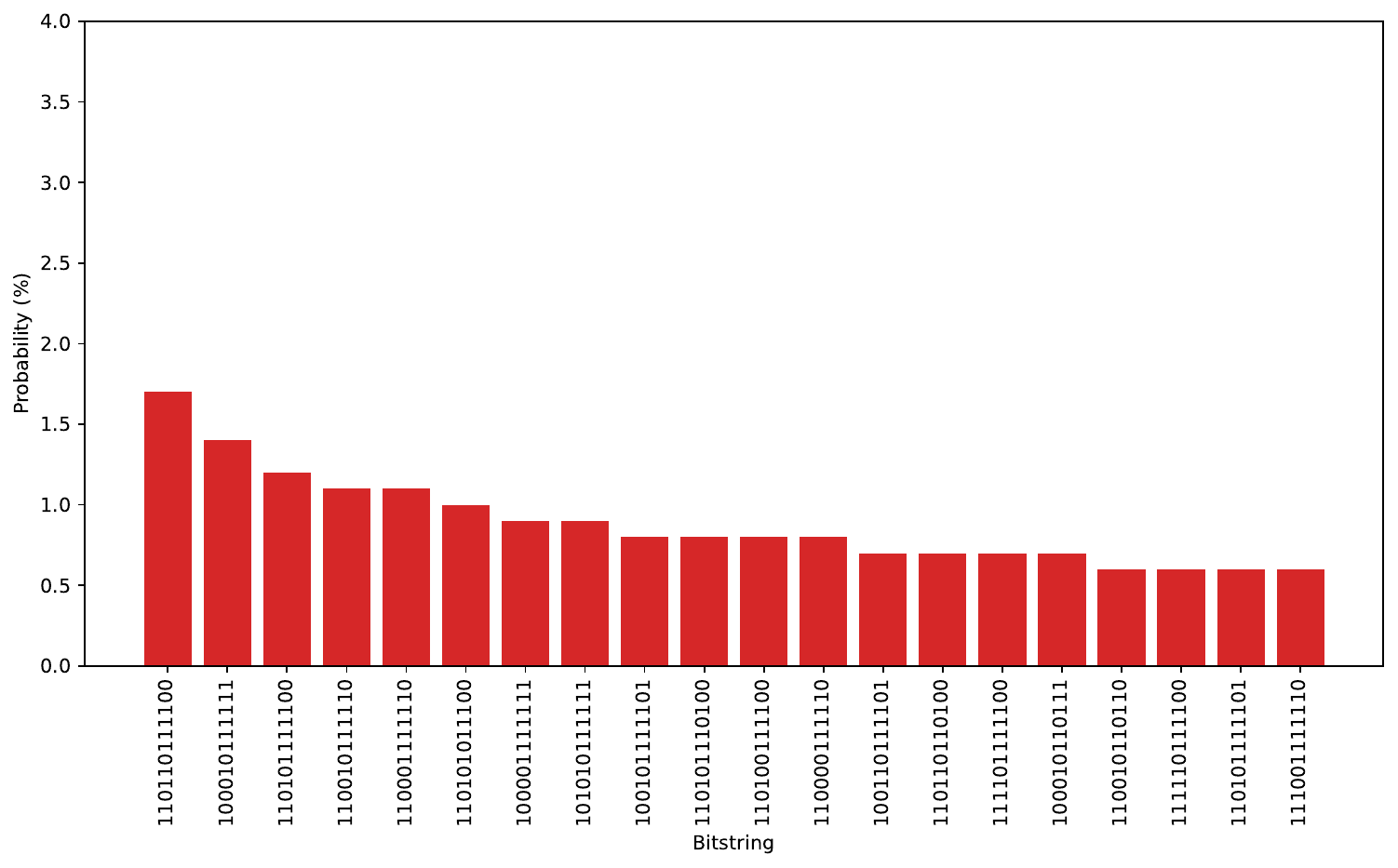}
    }
    \caption{Probability (in percentage) of reading a certain bistring for the experiments on Advantage QPU, both without and with the silhouette constraint. (a) shows the 20 most likely bitstrings from the experiment that does not use the silhouette information, while (b) the 20 most likely bitstrings from the experiment that uses the silhouette information as a constraint.}
    \label{fig:occurrences_dwave}
\end{figure}

\begin{table}[ht!]
    \centering
    \begin{tabular}{||l|c||}
        \hline
        \textbf{Platform} & \textbf{Total time} \\ 
        \hline 
        Fresnel (TN emulator) & $2.7 \cdot 10^{7} \mu s$ \\ 
        \hline 
        Fresnel & $1.00 \cdot 10^{5} \mu s$ \\
        \hline 
        Advantage QPU (without silhouette) & $8.82 \cdot 10^{4} \mu s$ \\
        \hline 
        Advantage QPU (with silhouette) & $1,30 \cdot 10^{5} \mu s$ \\
        \hline 
    \end{tabular}
    \caption{Comparison of the execution times for the clustering aggregation algorithms on the different quantum platforms.}
    \label{tab:time_comparisons}
\end{table}
Figure \ref{fig:occurrences_pasqal} shows the probabilities (in percentage) of reading a given bitstring from the output of the experiments on Fresnel (both emulated and real hardware), while Figure \ref{fig:occurrences_dwave} shows those related to both experiments on Advantage QPU.

\begin{figure}
    \centering
    \subfigure[Fresnel]{
        \includegraphics[width=0.75\linewidth]{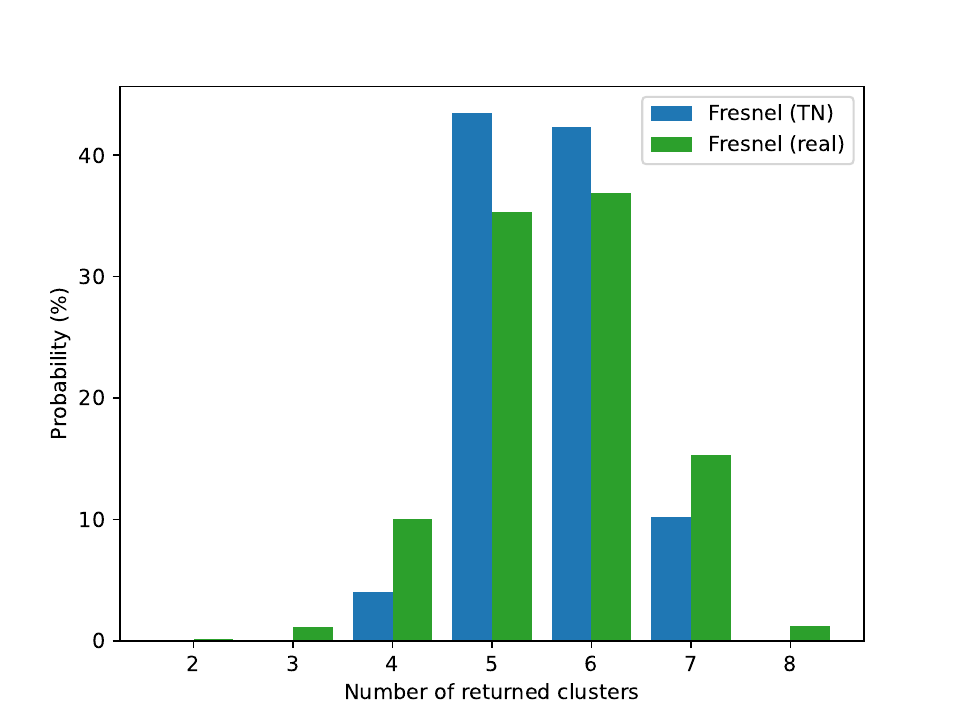}
    }
    \subfigure[Advantage QPU]{
        \includegraphics[width=0.75\linewidth]{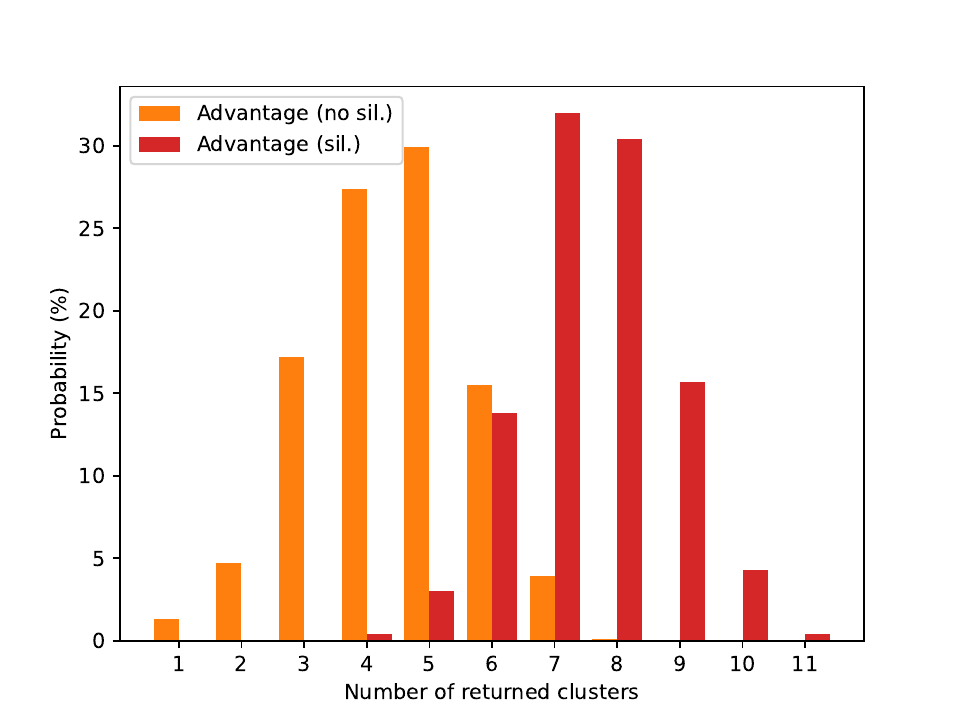}
    }
    \caption{Percentage of bitstrings returning a certain number of clusters fo the benchmark experiments on the various quantum platforms.}
    \label{fig:results_occurrences}
\end{figure}

Figure \ref{fig:results_occurrences} shows the probability of returning of a bistring with a given number of clusters for the experiments on the four different platforms. Results show how only the experiment on the Advantage QPU with the silhouette constraint added return the correct number of 7 clusters with the highest probability (around 30\%).

\section{Conclusion}

In this work, we proposed a hybrid algorithm to perform clustering aggregation, a technique in data science that aims to mitigate the issues of clustering algorithms (which tend to perform poorly on certain data distributions), by combining the results of multiple clustering algorithms in a more robust consensus clustering. The algorithm leverages the possibility of expressing clustering aggregation as a Maximum-Weight Independent Set (MWIS), where the ideal clustering must contain non-overlapping clusters that maximize the sum of weights attributed to them, in this case the sum of their silhouette scores. The next step was to partially reduce the MWIS to problems addressable by the chosen quantum machines, QUBO problems for the Dwave quantum annealer, and MIS problems for Pasqal's neutral-atom quantum computer.
Tests of the algorithm were performed using Pasqal's Fresnel, a neutral-atom-based quantum computer, and D-Wave's Advantage, a quantum annealer. A first assessment was performed on a smaller dataset and compared with the results obtained by employing a Fresnel emulator based on QuTiP. An experiment on a larger dataset was then performed to compare the results with Advantage QPU from Dwave, as well as with a Fresnel tensor network emulator accessible on Pasqal's cloud-based platform. The metrics chosen were the time required to complete the task and the probability of returning solutions with the correct number of clusters.
The results show that there is still much room for improvement; even in the experiment on Advantage QPU that included the silhouette information, only 30\% of the returned solution contained the correct number of clusters. At the same time, these results demonstrate that the design of hybrid algorithms that include neutral atoms and annealing technologies is a viable path to follow, and that the development of such algorithms could be used as a way to benchmark machines based on different technologies in terms of the quality of the results they produce.
Current technical limitation in Pasqal's Fresnel have shown of a higher degree of maturity of D-Wave's Advantage QPU; at the same time, the perspectives of register flexibility offered by a neutral-atoms based processor are interesting and represent a new technology worth exploring, since future technical advancements in neutral atom quantum computers, allowing to control each qubit individually, may lead to important refinements in future tests of the algorithm.
As future development of this work, it may be worth exploring the development of a similar algorithm that employs gate-based quantum computers, both to broaden the comparison of technologies and to investigate whether it is a more suitable class of solvers.
While it was not possible to demonstrate a quantum advantage either on the neutral-atom-based processor or on the quantum annealer, this work was able to corroborate the feasibility of addressing practical, industry-relevant problems, such as a clustering aggregation problem with quantum technologies, and represents another step towards the integration of quantum computers as accelerators for classical computers in hybrid quantum-classical pipelines.

\vspace{1.75cm}


\bibliographystyle{IEEEtran}
\bibliography{bibliography}

\end{document}